\documentclass[11pt,twoside]{book} 
\usepackage{asp2008n}
\usepackage{times}
\usepackage{lscape}
\usepackage{epsf}

\usepackage{graphicx}
\usepackage{amssymb}
\usepackage{longtable}
\usepackage[figuresright]{rotating}
\usepackage{multirow}

\newcommand{\simless}{\mathbin{\lower 3pt\hbox {$\rlap{\raise 5pt\hbox{$\char'074$}}\mathchar"7218$}}}

\mainmatter

\newlength{\deftabcolsep}
\begin{document}

% margin guide

\title{Where are all the Young Stars in Aquila?}
\author{L. Prato}
\affil{Lowell Observatory, 1400 West Mars Hill Road, Flagstaff, AZ 86001, USA}
\author{E.~L. Rice}
\affil{UCLA, Department of Physics \& Astronomy, UCLA,
Los Angeles, CA 90095-1547, USA}
\author{T.~M. Dame}
\affil{Harvard-Smithsonian Center for Astrophysics, 60 Garden Street,
Cambridge, MA 02138, USA}

\begin{abstract}
The high Galactic longitude end of the Aquila Rift
comprises the large Aquila molecular cloud complex, however,
few young stars are known to be located in the area,
and only one is directly associated with the Rift.  In contrast,
the Serpens star-forming region at the low Galactic longitude
end of the Rift contains hundreds of young stars.  We review
studies of the raw molecular material and describe
searches for young objects in the Aquila clouds.  The characteristics
of the known young stars and associated jets and outflows
are also provided.  Finally, we suggest some possible
explanations for the dearth of star formation in this gas-rich
region and propose some future observations to examine
this mystery further.

\end{abstract}

\vspace{-0.5in}

\section{Introduction}

 {The Aquila Rift} forms a great mass of dark clouds along the
summer Milky Way through the constellations Aquila, Serpens,
and eastern Ophiuchus.  Large scale plates several
degrees in diameter show almost a continuum of bright background
stars along the Galactic plane with dark patches of nearly
starless dark regions superimposed (e.g., Figure 1).  This structure is
silhouetted in the H$\alpha$ images of Madsen \& Reynolds (2005)
and reflected in the H~I map of Kawamura et al. (1999).
Curiously, in spite of the resemblance of these dark clouds to other
nearby, low-mass star-forming regions, few young stars have
been identified in the eastern (higher Galactic longitude) portion
 of the Rift. By contrast, the western portion of the {Aquila Rift}
contains the well-known Serpens star-forming region, near
Galactic longitude 30\,${\deg}$ but substantially above the
plane at a latitude of $\la$5\,${\deg}$.

\begin{figure}[t]
%\plotone{prato_fig1.eps}
\includegraphics[draft=False,width=\textwidth]{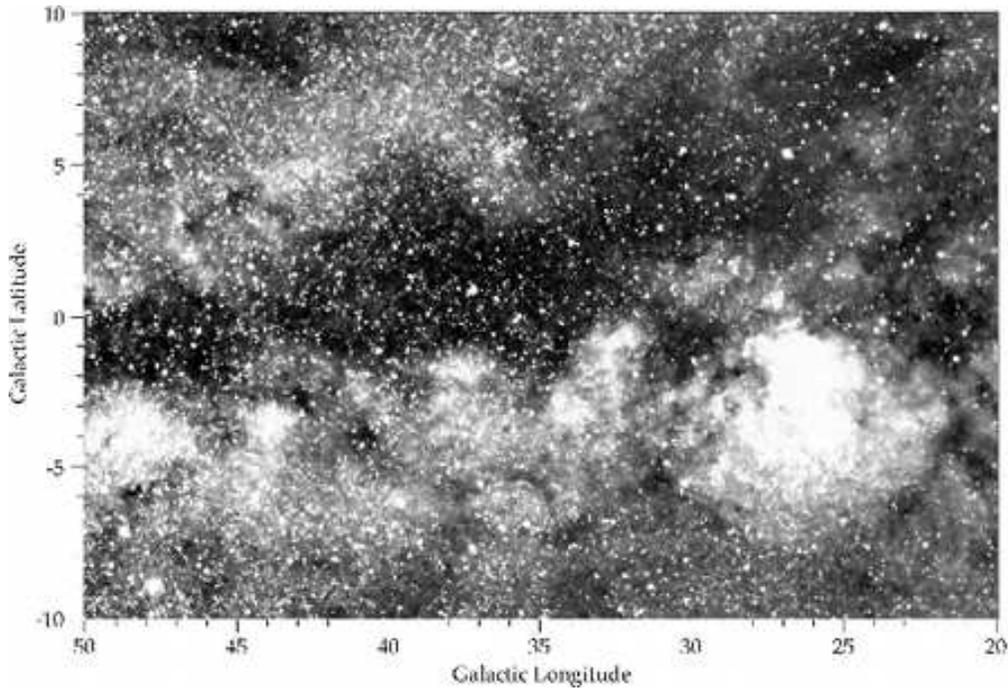}
\caption{A visible light panorama of a portion of the first Galactic
 quadrant of the Milky Way taken by A. Mellinger (de Cicco 1999). {The
 Aquila Rift} is the dark structure oriented diagonally across the image.}
\end{figure}

In this paper we briefly review the results of millimeter and submillimeter
surveys that include the eastern Aquila clouds (Section 2), describe the
interesting characteristics of the region's known young stars
(Section 3), discuss the best estimates for clouds' age, distance, and
relationship with the known young
stars in the area (Section 4), and speculate as to why much larger
numbers of young stellar objects are not known to populate the region
(Section 5).  In Section 6 we propose a number of potential observations
to determine better Aquila's star formation properties.
In this chapter we will focus on the high Galactic longitude Aquila
dark clouds. We will thus make a distinction between ``Aquila'', by
which we mean the region within a few degrees of the Galactic plane
located approximately between Galactic longitudes $30\,{\deg}$ and
 $50\,{\deg}$, and the larger {``Aquila Rift''} cloud structure (see Section 2).

\begin{figure}[t]
%\plotone{prato_fig2.eps}
%\epsscale{0.5}
\includegraphics[width=\textwidth,draft=False]{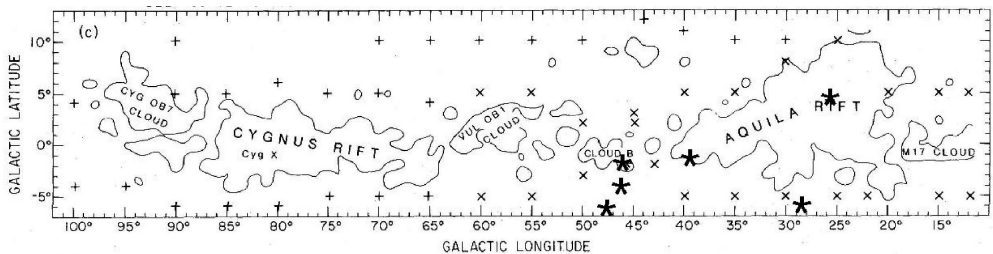}
\caption{Broad features of the $-$10 to $+$20~km~s\,$^{-1}$ CO
emission in the first Galactic quadrant.  The approximate
locations of the known young stars in Table~1 are indicated with asterisks.
The location of the early type Serpens star
 {MWC 297} ($l=26.80\,{\deg}$ and $b=+3.53\,{\deg}$) is also
marked.  Plot from Dame \& Thaddeus (1985).}
\end{figure}

\section{The Raw Materials:  Millimeter and Submillimeter Surveys}

The initial Galactic plane CO survey of Dame \& Thaddeus (1985) identified
 the salient features of the {Aquila Rift} and established a connection
between the dark nebulae and the molecular clouds in the region.
Subsequent improvements to the initial survey (Dame et al. 1987, 2001)
increased the sensitivity by more than a factor of 10, increased
the coverage of the CO observations from a 4\,${\deg}$ to a 10\,${\deg}$
strip in Galactic latitude, and increased the angular resolution
 to 1/8\,${\deg}$. In Galactic coordinates, the {Aquila Rift} stretches
from 20\,${\deg}$ to 40\,${\deg}$ in longitude and $-$1\,${\deg}$
to 10\,${\deg}$ in latitude, as demarcated by CO and 21~cm HI
(Dame et al. 2001; Figure 2).

 The molecular mass of the {Aquila Rift} as determined from
CO observations has been estimated to be between
1.1$\times10\,^5$~M$_{\odot}$ and 2.7$\times10\,^5$~M$_{\odot}$
(Dame \& Thaddeus 1985; Dame et al. 1987; Strai\v{z}ys et al. 2003).
Using the molecular line widths and a uniform density sphere
approximation, Dame \& Thaddeus
 calculate the virial mass of the {Aquila Rift} to be
2.6$\times10\,^5$~M$_{\odot}$. The virial mass is then about the same
as or slightly greater than the observed mass, depending on the
parameters used in the H\,$_2$ gas mass estimates, particularly the cloud
distance and therefore size and density.  The relationship
between the virial mass and the observed mass reflects
the star-forming potential of these
clouds because, if the virial mass dominates, the cloud is
dynamically unstable and unlikely to form stars (e.g., Solomon et al. 1987).
 Therefore, the ambiguity in the measured mass of the {Aquila Rift}
is important to follow up with
additional millimeter observations and improved distance measurements.

 {The Aquila Rift} consists of numerous small and large clouds, many of
which have been identified and tabulated by Lynds (1962). Unfortunately,
the Lynds coordinates are sometimes so uncertain that it is not clear
which cloud they refer to. Dobashi et al. (2005) have
performed an extinction study of the Galactic plane using automated star
counts, and they offer a list of clouds with finding charts and accurate
coordinates.

Kawamura et al. (1999) used $^{12}$CO observations to search for
 molecular clouds in the region to the Galactic south of the {Aquila Rift.}
Although they identified dozens of small clouds, possibly
dynamically connected to the Rift, no correlation
with stellar $IRAS$ point sources was apparent.  Kawamura et al.
(2001) focussed on the region in the immediate vicinity of the
 T~Tauri star {HBC~294} {(V536~Aql).} They detected a ring shaped
cloud in $^{12}$CO ($l=48.1\,{\deg}$, $b=-6.3\,{\deg}$),
also seen in the Dame et al. (2001) survey,
containing about 430~M$_{\odot}$ of gas,
five $^{13}$CO cores, and three C$^{18}$O cores.  However, no
candidate young stellar objects, as indicated by $IRAS$ fluxes,
were found.
Complementary objective prism observations did detect an
H$\alpha$ emission line object, which may be a young star, near
 the dark cloud {LDN~694} (Kawamura et al. 2001).

Harvey et al. (2003) studied the properties of the
 candidate protostellar collapse core {Barnard 335} in millimeter
continuum and compared these data with prior observations of the core
in CS (Wilner et al. 2000) and in NH\,$_3$ (Benson \& Myers 1989;
see also the chapter by Reipurth on Bok globules).
This core is located only a few degrees from the molecular ring seen by
 Kawamura et al. (2001) and is associated with the dark cloud {LDN 663}
(see Figure 6 in Kawamura et al. 2001).

Additional millimeter and submillimeter surveys have been conducted,
using both line and continuum emission, to search for signs of dense
 molecular and possibly pre-stellar cores in the {Aquila Rift} region.
Anglada et al. (1997) focussed on NH\,$_3$ in areas where outflows had
been previously detected in optical light and in molecular lines.
Morata et al. (1997) observed CS in the region near the outflows in
the filamentary dark cloud, about 15\,$'$ north of the prominent
 T~Tauri star {AS 353} (Section 3.7). Their strongest detection of CS
 coincides with the location of the dark cloud, {LDN~673} (Lynds 1962;
Figure 3).  Morata et al. (2003, 2005) analyzed the small-size
structure of this region in a multitransitional study including
interferometric maps.  Kirk et al. (2005) used submillimeter JCMT
observations to study five dark clouds in the area; no pre-stellar
cores were found in the four Lynds clouds in the sample.  However, the
 object {Barnard 133,} located within about a degree of several known
 young stars just off the southern Galactic edge of the {Aquila Rift,} is
not only a strong submillimeter source but also harbors an NH\,$_3$
core (Benson \& Myers 1989).  Visser et al. (2002) performed a large
 850~$\mu$m survey of {LDN~673} and found eight sources, SMM~1-8, some of
which are associated with IRAS sources (see their Figure~20).  Many
ultra-compact HII regions, tracers of high mass star formation, are
 located throughout the {Aquila Rift} (Becker et al. 1994), however,
their distances are ambiguous and the majority are likely to be
background objects.

Abundant raw material appears to be available
in the region, although few cores harboring active
star formation have been identified.  The few known young stars
(Section 3) and star-forming cores are almost all scattered throughout the small
 molecular cloud clumps to the Galactic south of the {Aquila Rift.}
The eastern portion of the Rift is apparently rich in gas but almost
devoid of known star formation (Figure 4).

\begin{figure}[!ht]
%\plotfiddle{prato_fig3.eps}{12cm}{0.0}{28.5}{28.5}{-185.0}{0.0}
\includegraphics[width=\textwidth,draft=False]{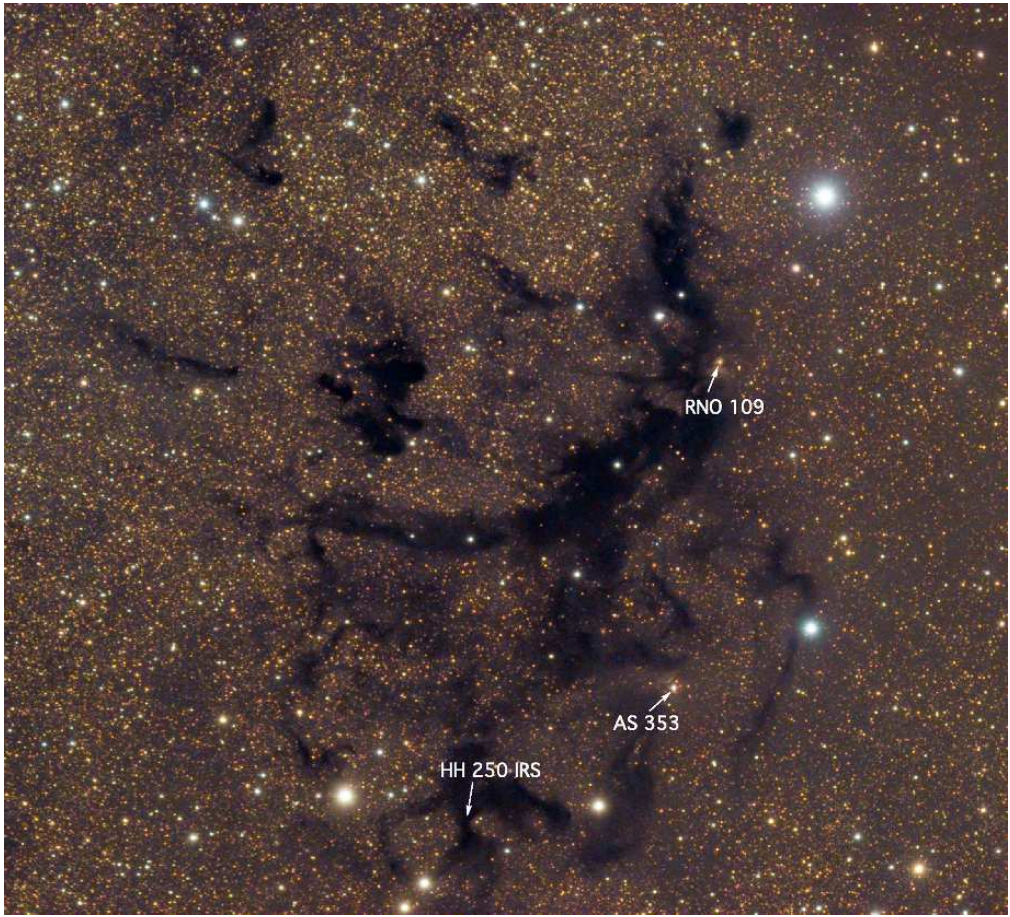}
 \caption{The {LDN~673} cloud complex is highly fractured. The image is
approximately 1 degree wide; north is up and east to the
left. Courtesy of Bernhard Hubl.}
\end{figure}

\begin{figure}[t]
%\plotone{prato_fig4.eps}
\centering
\includegraphics[width=\textwidth,draft=False]{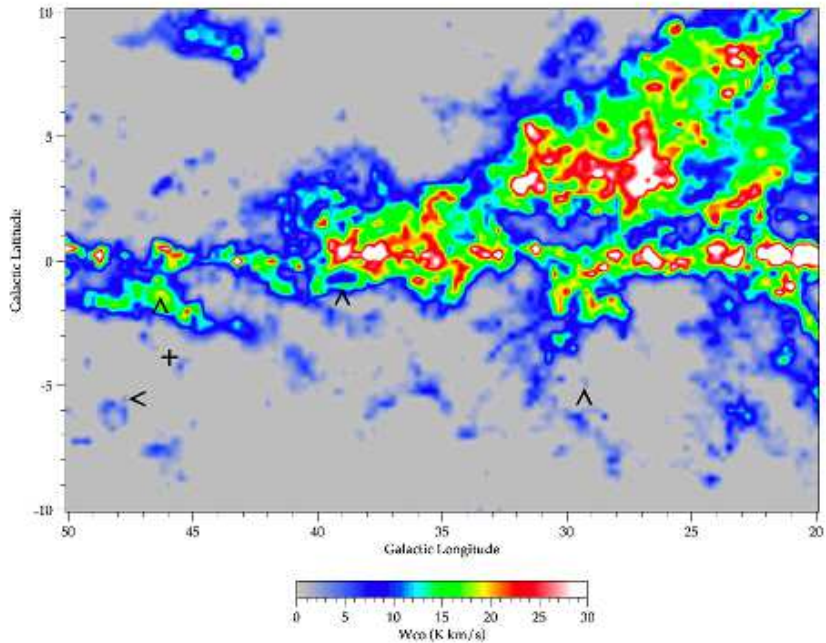}
 \caption{An integrated $^{12}$CO map of the {Aquila Rift,} from
$v_{LSR}=-30$ to $v_{LSR}=+30$~km~s\,$^{-1}$.  The high Galactic longitude
sector of the Rift, from $l$ of 33${\deg}$ to 41${\deg}$
and $b$ of 3${\deg}$ to $-$2${\deg}$ only contains one known
 young star, the unusual object {HBC 684,} on the lower extreme of
the cloud.  The Serpens star-forming region is located
around $l\sim30{\deg}$ and
$b\sim5{\deg}$.  Arrows indicate molecular material associated
with known young stars (Table 1).  A cross shows the position of
 {Parsamian 21.}}
\end{figure}

\section{The Known Stars: Visible and Infrared Studies}

Several young objects in Aquila have been very
well studied (Section 3.7), but the remainder of the known young stars have
not received much attention until recently, although Cohen \& Kuhi
(1979) identified a few of them.  Indeed, it is not
yet known what the total stellar census may be in the region.
Th\'e (1962) found several H$\alpha$ emission-line objects towards Aquila
and Scutum, however, followup spectroscopy is required to identify
the nature of these detections.
The known young Aquila stars are listed in Table 1 and their
approximate positions illustrated in Figures 3 and 4.

\begin{table}[b]
\caption{Known Young Aquila Stars}
{\small
\begin{tabular}{lcccccc}
\multicolumn{1}{l}{Object} &
\multicolumn{1}{c}{HBC} &
\multicolumn{2}{c}{RA (J2000) DEC} &
\multicolumn{1}{c}{$l, b$ (${\deg}$, ${\deg}$)} &
\multicolumn{1}{c}{SpTy} &
\multicolumn{1}{c}{V mag} \\
%\hline
\tableline
\noalign{\smallskip}
FG~Aql/G1 & 681 & 19 02 22.2 & $-$05 36 20 & 29.17, $-$4.98 & K5 & 13.7 \\
FG~Aql/G3 & $...$ & 19 02 22.6 & $-$05 36 22 & 29.17, $-$4.98 & M0 & 15.5 \\
FG~Aql/G2 & 682 & 19 02 22.8 & $-$05 36 15 & 29.18, $-$4.98 & K5 & 13.6 \\
FH Aql & 683 & 19 02 23.2 & $-$05 36 37 & 29.17, $-$4.98 & K7 & 15.5 \\
{\small {\it IRAS}19046$+$0508} & 684 & 19 07 09.8 & $+$05 13 10 & 39.37, $-$1.11 & K5 & 15.5 \\
AS~353A  & 292 & 19 20 31.0  & $+$11 01 54 & 46.05, $-$1.33 & K5 & 12.5 \\
AS~353B & 685 & 19 20 31.0 & $+$11 01 49 & 46.05, $-$1.33 & M0 & 14.6 \\
Parsamian 21 & 687 & 19 29 00.7 & $+$09 38 39 & 45.82, $-$3.83 & F5 & 14.2 \\
V536 Aql & 294 & 19 38 57.4 & $+$10 30 16 & 47.75, $-$5.57 & K7 & 14.9 \\
\tableline
\end{tabular}
}
\end{table}

Rice et al. (2006) have lately completed a high spectral resolution
study in the infrared of all but one of the Aquila stars
listed in the Herbig and Bell Catalogue (Herbig \& Bell 1988; HBC).
 The exception, {Parsamian 21,} which has been described as a
candidate \normalsize{FU~Ori} object (e.g., Staude \& Neckel 1992), is the only
potential higher-mass object in Aquila and may in fact actually be
a background object at $\sim$2 kpc (Section 3.7).  The initial goal
of Rice et al. was to determine the radial velocities
of the sample and to search for variability indicative of spectroscopic
binaries.  The study
ultimately revealed a wealth of information about the low-mass targets.

\subsection{Stellar Radial Velocities}

Radial velocities can be used to strengthen the association among
young stars in close spatial proximity on the sky.  Rice et al. (2006)
used high-resolution (R$\sim$30,000) $H$-band spectra of
eight of the nine objects listed in Table 1 in order to determine
radial velocities, rotational velocities, and spectral types, as well
as to search for radial velocity variability that would indicate an
angularly unresolved
companion.  The precision of the radial velocity measurements
was estimated to be 2~km~s\,$^{-1}$.  The radial velocities were
measured from high signal-to-noise ratio spectra (SNR$\sim$200$-$400)
and from spectra typically obtained at multiple
epochs.  Although the young stars in Aquila
are spread out over $\sim20\,{\deg}$ on the sky, the dispersion of
the radial velocity measurements made by Rice et al. (2006) is about
2~km~s\,$^{-1}$, suggesting that the stars formed from the same
molecular cloud complex (e.g., Herbig 1977).
 The similar radial velocities of {AS~353A} (average of
 $-$11.4~km~s\,$^{-1}$ from three epochs) and {AS~353B}
($-$10.7~km~s\,$^{-1}$, from one epoch), along with the common proper
motions measured by Herbig \& Jones (1983), indicate that this system
is physically related.  The average radial velocity for eight of the
 objects studied was $-$8.6~km~s\,$^{-1}$; a ninth object, {HBC 682,} was
significantly variable in radial velocity and is hence a spectroscopic
binary candidate.  The $v{\sin}i$ rotational velocities of this
sample measured by Rice et al. (2006) range from 10~km~s\,$^{-1}$ to
50~km~s\,$^{-1}$.

\subsection{X-Ray Properties}

X-ray luminosity in young stars is thought to be thermal emission from gas
heated by magnetic reconnection events between the magnetic field of
the star and that of the circumstellar disk.  Surveys of nearby
star-forming regions reveal dozens or hundreds of X-ray sources,
associated mainly with weak-line T~Tauri stars (Feigelson \& Montmerle
1999 and references therein).  Queries of the \emph{ROSAT}, Chandra,
and XMM/Newton archives using HEASARC
reveal only two sources within 5 arcminutes of any
of the coordinates listed in Table~1.  Both of these sources were
found in the \emph{ROSAT} All-Sky Survey (RASS) Faint
Source Catalog (Voges et al. 2000).  All RASS
sources were observed for an average of $\sim$500 seconds and with a
detection limit of 6 photons (Belloni et al. 1994).  One detected
 source near the young stars in Aquila lies within 1 arcminute of {FG~Aql/G1,}
 {FG~Aql/G2,} {FG~Aql/G3,}
 and {FH~Aql,} and the second source is about 3 arcminutes from {HBC~684.}
Both of these sources have hardness ratios consistent with that of
known T~Tauri stars (e.g. Neuh\"auser et al. 1995, Kastner et
al. 2003).  The only Chandra observations near any of the Table~1
objects are more than 30 arcminutes away, and the nearest
XMM/Newton observation is more than 70 arcminutes away.

\subsection{Extinction}

Typical stellar extinctions found by Rice et al. (2006)
range from $A\,_v\sim0-4$ magnitudes.  Based on the \emph{2MASS} JHK
magnitudes, these represent interstellar extinction to the scattering
surface of the stellar system (i.e. circumstellar disks and shells) and
may underestimate the total extinction to the stellar photosphere
in the presence of circumstellar material (e.g., Section 4.2, Prato et al.
2003).  In general, the values of interstellar extinction to the Aquila
young stars are consistent with the typical values found for the
region's molecular clouds, suggesting a similar distance.  The dust maps of
Schlegel et al. (1998) indicate an upper limit on the extinction
 in the {Aquila Rift} of $A\,_v\sim5$ magnitudes (Drew et al. 2005).
 Dobashi et al. (2005) find maximum extinctions along the {Aquila Rift}
of 5$-$10 magnitudes.  The
photometric study of $\sim$500 stars by Strai\v{z}ys et al. (2003)
suggests a maximum $A\,_v$ of about 3.0 magnitudes throughout the Rift and
a distance to the front edge of the clouds of 225 $\pm$55 pc (Section 4.1).

\subsection{Multiplicity}

The multiplicity among the few known young Aquila stars
appears comparable to that of Taurus, the nearby star-forming region with
the highest binary fraction (e.g., Ghez et al. 1993; Simon et al. 1995).
For the nine systems listed in Table~1, there are a total of at least
14 primary and companion objects.
 {HBC~681} and {HBC~682} were identified as
visual binaries by Rice et al. (2006) for the first time.  HBC~682A
is also a candidate spectroscopic binary (Rice et al.).
The {HBC~294} system
 is a known, subarcsecond binary (Ageorges et al. 1994) and {AS~353}
a known hierarchical triple (Tokunaga et al. 2004).

\subsection{Circumstellar Disks}

Most of the objects in Table 1 lie in the region of the J$-$H vs. H$-$K
color-color diagram characterized by an IR excess (Figure 5), indicating
the likelihood of abundant circumstellar disk material in these systems.
 In particular, Figure 5 shows that {AS~353A} and {HBC~687} {(Parsamian 21)}
both lie {\it below} the classical T~Tauri star locus.  This may be
an indication of abundant reflected light from circumstellar material.
 {Parsamian 21} is probably a background \normalsize{FU Ori} type object
with an unusually
massive disk (Staude \& Neckel 1992).  The circumstellar disk of
 {AS~353A} has been well-studied; this system is one of the most active
and visually bright T~Tauri stars known (e.g., Tokunaga et al. 2004
 and references therein). {HBC~684} is a unique emission line
object, possibly surrounded by a massive disk.  It is by far the
 reddest object in Table 1. {FG~Aql/G1,} {FG~Aql/G2,} and {V536~Aql} all
exhibit the ordinary behavior of classical T~Tauri stars with ongoing
accretion from a circumstellar disk.  In summary, like the
multiplicity fraction, the circumstellar disk
fraction of this group of stars is relatively high compared to other
nearby star-forming regions.  Several of these systems are discussed in
detail in Section 3.7.

\begin{figure}[!hb]
%\plotfiddle{prato_fig5.eps}{11.5cm}{90}{67}{67}{310}{-35}
\centering
\includegraphics[width=0.75\textwidth,draft=False]{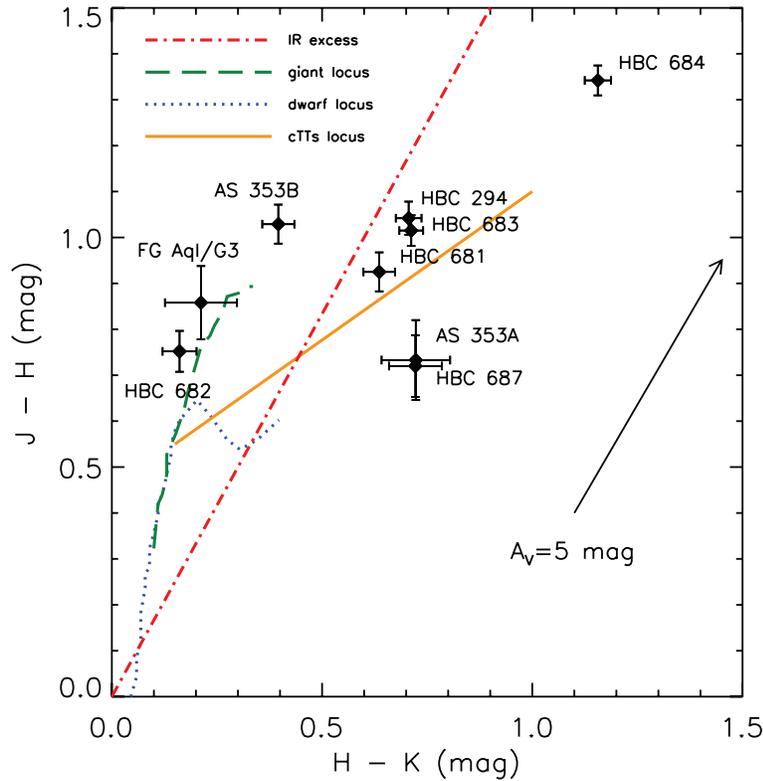}
\caption{$J-H$ versus $H-K$ color-color diagram for the objects
in Table~1. Magnitudes are from \emph{2MASS} and the error bars
represent the propagated photometric uncertainties reported
in the that catalog.  The dash-dot line separates objects
with (to the right) and without IR-excess.  The cTTs (solid line), dwarf
(dotted line), and giant (dashed line) loci are the same as in Figure~4
of Prato et al. 2003 but transformed in to the \emph{2MASS} magnitude scale
using equations from Carpenter 2001.  The effect on observed
color of 5 magnitudes of visual extinction is represented by the
arrow (thick line), using the equation derived by Prato et al. 2003.}
\end{figure}

\subsection{Herbig-Haro Jets and Outflows}

Herbig-Haro flows are signposts of recent star formation, and they
are thus of great interest in identifying very young stars.
A detailed discussion of HH flows and their energy sources can
be found in Reipurth \& Bally (2001). While the Aquila clouds are
too large to have been fully surveyed by CCD images, the Palomar
Schmidt plates have been examined and CCD images have been obtained
of regions around the known young stars. As a result, a number of HH
objects are known in Aquila, and they are discussed individually below.

\subsubsection{HH 32 from AS 353A}

 {HH 32} is a bright HH object originating in {AS 353A;} it was discovered by
 Herbig (1974). It consists of two main bow shocks, HH 32A and 32B. {HH 32}
is a high-excitation HH object for which optical spectroscopy has been
reported by Dopita (1978), Brugel, B\"ohm, \& Mannery (1981a,b),
Herbig \& Jones (1983), Solf, B\"ohm, \& Raga (1986), and Hartigan,
Mundt, \& Stocke (1986). Ultraviolet spectroscopy was reported by
 B\"ohm \& B\"ohm-Vitense (1984) and Lee et al. (1988). {HH 32} is one
of the rather rare red-shifted HH objects, and the radial velocity
measurements, combined with proper motions determined by Herbig \&
 Jones (1983) and Curiel et al. (1997), show that it moves away from {AS
 353A} with a space velocity of about 300~km~s\,$^{-1}$ at an angle of about
70\,${\deg}$ to the plane of the sky.
The fine details of this HH flow are seen in the HST images of
Curiel et al. (1997; Figure 6).  Beck et al. (2004) obtained an
 integral field unit data cube for {HH 32,} providing the hitherto most
detailed spectroscopic and kinematic study of this HH flow.  These
data were modelled by Raga et al. (2004) in terms of the internal
working surface model.  Mundt, Stocke, \& Stockman (1983)
 found a third flow component, HH 32C, on the opposite side of {AS 353A}
 with high blue-shifted velocities, indicating that {HH 32} is a bipolar
flow.

Edwards \& Snell (1982) observed
 high-velocity CO emission associated with {AS 353A} and the {HH 32} flow.
 {HH 32} is one of the few HH objects detected in the radio continuum
(Anglada et al. 1992, 1998).  It also emits in the infrared H\,$_2$ lines
(Davis, Eisl\"offel, \& Smith 1996, and references therein).
 Davis et al. (1996) report three faint HH knots, {HH 332,} about an
 arcminute south-west of {HH 32,} but not on the well-defined flow axis
 of {HH 32.} Tokunaga et al. (2004) argue that these are likely to
 be part of an earlier precessing flow component of {HH 32,} since the
flow is significantly foreshortened, so a small angle in flow
direction projects to a much larger angle on the sky.

\begin{figure}[!hbt]
%\plotfiddle{prato_fig6.eps}{9.5cm}{-90.0}{55.0}{55.0}{-230.0}{295.0}
\centering
\includegraphics[width=0.7\textwidth,draft=False]{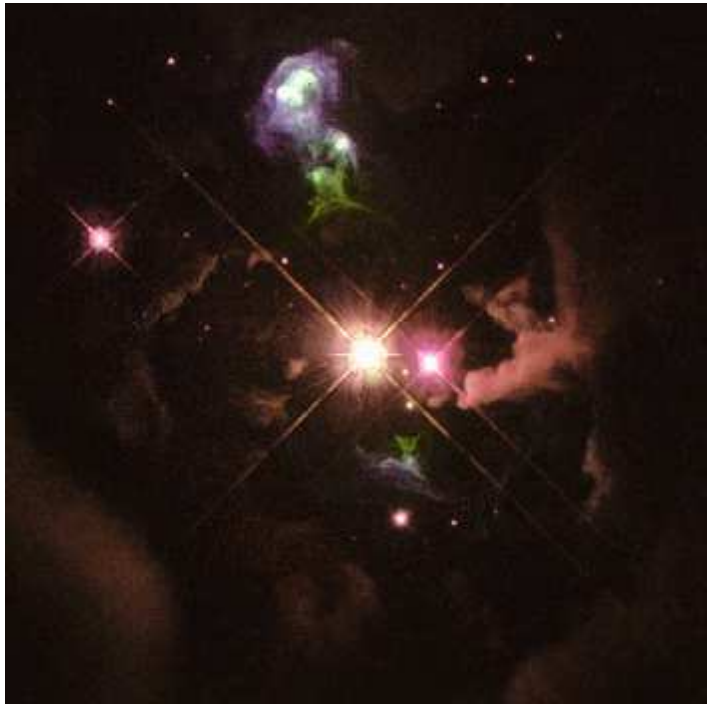}
 \caption{AS 353A and B together with the bipolar {HH 32} flow as seen
  with the HST. The young stars have created a wind-blown cavity in
  their natal cloud, seen in reflected light. Composite image from R,
  H$\alpha$, [NII], and [SII] images. From Curiel et al. (1997).}
\end{figure}

\subsubsection{HH 250 from IRAS 19190+1048}

 About 14 arcminutes southeast of {HH 32,} Devine, Reipurth, \& Bally
(1997) discovered a rather bright compact bow shock facing away from
 the embedded {IRAS source 19190+1048.} Another fainter HH knot is found
closer to the source, and two diffuse reflection nebulae surround the
source. The energy distribution of the IRAS source suggests that it is
a Class~I object.

\subsubsection{RNO 109}

 Within the cloud {LDN 673,} Armstrong \& Winnewisser (1989) reported the
discovery of a molecular outflow associated with a small nebulosity,
 {RNO 109,} a feature associated with the IRAS point source 19180$+$1116
(Cohen 1980) and having colors characteristic of a probable embedded
young stellar object.  Coordinates for this object are poorly defined
and no spectra are available in the literature.  Thus, we do not
 include {RNO 109} in Table 1. Another nearby source, {IRAS 19180$+$1114,}
also drives an outflow (Visser et al. 2002).  This region is only a
 few degrees from the {AS 353} system and located within the same
molecular cloud condensation (Figure 3).  The outflow was observed in
CO and $^{13}$CO and extends over 10$-$15 square arcminutes.  Morata
et al. (1997) observed this outflow in CS gas, and Anglada et
al. (1997) in NH\,$_3$.

\subsubsection{HH 119}

 The Bok globule {B335} is associated with HH object 119
(e.g., Reipurth et al. 1992).
This system is discussed in depth in the chapter
on Bok globules by Reipurth.

\subsubsection{HH 387}

 Hirth et al. (1997) first identified {HH 387,} associated with
 the subarcsecond binary {V536 Aql} (Section 3.7), as a small jet with a
3$-$4 arcsecond extent and PA of 90\,${\deg}$.  Mundt \& Eisl\"offel (1998)
confirmed their findings and, imaging the system in [SII],
identified knots in the jet as far as 16 arcseconds away at a PA of
110\,${\deg}$.  A very faint counter jet is suggested in the data
of Mundt \& Eisl\"offel that requires further confirmation.

\subsubsection{HH 221}

A very small jet-like feature is present along the polar
 axis of the conical nebula associated with {Parsamian 21}
(Staude \& Neckel 1992).  However, it is unlikely that this object
is a true Aquila member and is more probably at a
distance of $\sim$2~kpc.

\subsection{Notes on Individual Stars}

\subsubsection{AS 353}

 By far the best known young star in the Aquila region is {AS 353} (Figure 6),
originally discovered by Merrill \& Burwell (1950), and later
independently in the survey of Iriarte \& Chavira (1956).  The
 associated outflow, {HH 32,} is discussed in Section 3.6.
 Given the brightness (V$\sim$12.5) of {AS 353,} it can be studied in great
detail. Herbig \& Jones
(1983) provided the first detailed discussion of the emission line
 spectrum of {AS 353A,} and Mundt, Stocke, \& Stockman (1983) presented
high-resolution H$\alpha$ and sodium doublet spectra, documenting the
massive, high-velocity, neutral wind of the star. Further echelle
spectra demonstrated that the H$\alpha$ profile shows significant
 variability (Hartigan, Mundt \& Stocke 1986). In their study of {AS 353A,}
B\"ohm \& Raga (1987) presented spectrophotometry from 3,300 \AA\ to
10,000 \AA\ and noted the strong ultraviolet excess of the star. In a
subsequent and more detailed study, Eisl\"offel, Solf, \& B\"ohm (1990)
presented fluxes of all emission lines in about the same spectral
 range but with higher spectral resolution. {AS 353A} has been part of
numerous optical spectral studies of T Tauri stars since then,
including Hamann \& Persson (1992), Hamann (1994), and Edwards et
 al. (1994). In the near-infrared region, {AS 353A} shows the 2.3 $\mu$m
CO band in emission (Carr 1989; Biscaya et al. 1997; Prato et al. 2003;
Rice et al. 2006) as well as other emission lines (e.g., Davis
et al. 2003; Prato et al. 2003).  The stellar continuum is so
heavily veiled that a photospheric spectrum cannot be easily seen,
although Basri \& Batalha (1990) suggested a spectral type of K2.  Recent
multi-epoch high-resolution $H$-band spectra revealed the stellar
photospheric features of a K5 spectral type star (Rice et al. 2006).

 {AS 353A} displays considerable photometric variability
(e.g., Fern\'andez \& Eiroa 1996).  Its near- and mid-infrared photometry
is summarized by Rice et al. (2006) and Molinari, Liseau, \& Lorenzetti
(1993).  IRAS observations are discussed by Cohen \& Schwartz (1987),
sub-millimeter observations are reported in Reipurth et al. (1993),
and centimeter data are given by Anglada et al. (1998).
 {AS 353A} forms a triple system with the less well-studied {AS 353B,} a
weak-line T Tauri star binary of separation 0.24 arcseconds (Tokunaga et al. 2004).

\subsubsection{HBC 684}

 {HBC 684} is a spectacularly peculiar object. No photospheric absorption lines
were detected even in high-dispersion $H$- and $K$-band spectra, but
it was possible to measure a radial velocity, consistent with the other
Aquila young star velocities, using the pure atomic metal emission
spectrum observed in this object in three $H$-band epochs
spanning more than a year (Rice et al. 2006).  An emission line
spectrum was also observed in visible light two years after the
infrared observations, revealing a broad P~Cygni H$\alpha$ profile
with a $\sim$200~km~s\,$^{-1}$ absorption trough.
 {HBC 684} is the most extinguished Aquila object known and shows one of the
strongest infrared excesses (Figure 5); it is coincident with the
 {IRAS source 19046+0508.} The line ratios {\it in emission} are
consistent with a K5 spectral type object.  Rice et al. (2006) provide a
detailed discussion of this object and speculate as to the origin of
its unusual spectrum.

\subsubsection{HBC 294}

 Ageorges et al. (1994) first discovered {HBC 294} to be a binary with an
0\,\farcs52 separation at a position angle of 17\,${\deg}$.
 {HBC~294} is a typical, classical T Tauri
star, with Br$\gamma$ and Pa$\beta$ emission detected in near-infrared
spectra, H$\alpha$ and forbidden
emission lines seen in visible light, and a near-infrared excess, all
indicative of accretion from a circumstellar disk and disk-driven
outflows associated with at least one of the stars (Rice et al. 2006;
 Hamann 1994). The \emph{IRAS} PSC flux measurements of {HBC~294}
 {(IRAS~19365+1023)} reveal even greater mid- to far-IR fluxes
 than those of {AS~353A} at all but 100~$\mu$m (Weaver \& Jones 1992).
The lower 100~$\mu$m flux suggests
that the disk(s) may be truncated by the binary.

\subsubsection{Parsamian 21}

 {Parsamian 21,} or {Par 21,} was initially identified by Parsamian (1965)
from the Palomar Sky Survey plates during a search for cometary
nebulae.  Dibai (1969) identified the associated star as type A5V, and
Cohen (1974) observed the system in the mid-infrared and found $N$-
and $Q$-band magnitudes of 3.7 and 1.4, respectively.  It is
 associated with the source {IRAS 19266+0932.} Digital sky survey plates
show an unusual and complex nebulosity.  If this represents half of a
conical, bipolar structure, a high degree of obscuration must block
the unseen portion of the nebula.  Neckel \& Staude (1984) concurred
with the A5V spectral type designation, however, in a later paper
(Staude \& Neckel 1992), they identified \normalsize{FU Orionis}
 characteristics in
 the spectrum of {Par 21} and re-classified the type as F5Iab. The
visible light spectrum of the photosphere is characterized by broad
absorption lines and in particular a prominent P~Cygni profile with an
absorption trough of width $\sim$450~km~s\,$^{-1}$ (Greene, Aspin, \&
Reipurth 2008).  Evidence in the form of 1.3 mm emission (Henning et
al. 1998), prominent disk-pattern polarization (Draper et al. 1985),
strong near- and mid-infrared excesses (e.g., Cohen 1974), and water
ice absorption and crystalline silicate emission (Polomski et
al. 2005) point to the presence of a massive circumstellar disk.
Given the luminosity of this system, it is unlikely to be located in
Aquila and is most probably at a distance of 1.8~kpc (Staude \& Neckel
1992).

\section{Putting It All Together}

\subsection{Distance}

 For the {AS 353} system in Aquila, Prato et al. (2003) adopted a distance
estimate of 150 $\pm$50 pc.  Earlier estimates for the distance
 to the {Aquila Rift} range from 110~pc (Weaver 1949) to
150~pc (Edwards \& Snell 1982) to 200~pc
(Dame \& Thaddeus 1985).  Herbig \& Jones (1983) estimated the
 distance to {LDN~673} to be 300~pc and noted that {LDN~673} is foreground
to ``the very extensive Aquila obscuration.''  If the Aquila young
 stars are indeed associated with the {Aquila Rift} and are therefore
connected with the Serpens region, then the
distance of Serpens is also germane.  Unfortunately, this number
also has a history of uncertainty.  Racine (1968) and
Strom et al. (1974) estimated a distance of 440 pc to Serpens
 through studies of the star {HD 170634.} Both groups determined
an early spectral type, B7V and A0V, respectively, and applied a small
reddening correction, based respectively on visible and infrared light, to
derive the distance modulus.   De Lara et al. (1991) combined infrared
photometry and visible light spectroscopy of five stars in
Serpens, yielding an improved average distance modulus and
a distance of 311 pc.  Recently, Strai\v{z}ys et al. (2003), using
two-dimensional photometric classification of stars in the seven-color
Vilnius system, estimated a distance of
225 $\pm$55 pc.  Eiroa et al., in the chapter on Serpens,
adopt a distance of 230~$\pm$20~pc to the Serpens region.
This appears to be based on a rough average of recently derived values,
including unpublished estimates from 2MASS data and associated
extinctions.  We note that recent distance determinations
to Serpens are not only dropping, but are also converging with
estimates for the Rift in general, and for the Aquila young stars
specifically.  For now we adopt a distance  of 200 $\pm$30 pc
to Aquila, as in Rice et al. (2006), and stress the importance of
future observations to determine this quantity accurately (Section 6).

\subsection{Age}

A number of indications point to a young age for the Aquila
stars, including the high circumstellar disk fraction,
the striking Herbig-Haro jets driven by several
energetic sources, and the possible association with the Serpens
cluster (age $\sim$1 Myr; Winston et al. 2005).
Based on the evolutionary models of Palla \& Stahler (1999),
Prato et al. (2003) and
Rice et al. (2006) determined ages for the known young stars
 (Section 3) of at most a few Myr, with the exception of {FG Aql/G3}
which may be older than 10 Myr.  Ages derived for this
sample from the models of Baraffe et al. (1998) are similar.
Although these stars span a broad area on the sky, their approximately
common ages and similar radial
velocities suggest origins in the same cloud complex.

\subsection{Star Formation in Aquila}

Molecular maps of the Galactic plane region show that there is at
least low-intensity CO emission at the location of all the known young
Aquila stars (Figure 4).  However,
they are spread out over a large
area (30$\times$60~pc, assuming d$=$200~pc).  The CO with
which they appear to be associated is clumpy and filamentary.  The
stars are all located below the Galactic plane while the bulk of gas
present in the Aquila cloud extends along and to the north of the Galactic
plane.  Furthermore, there exists a complex velocity structure within
the Rift; a velocity
break in the CO at about  $l=33\,\deg$ appears to indicate a demarcation
between two distinct sets of clouds (Figure 7).  This
break might account for the
double-peaked CO lines seen by White, Casali, \& Eiroa (1995).

As shown by Rice et al. (2006; Section 3.1), the rms of the Aquila young stars'
radial velocities is only $\sim$2~km~s\,$^{-1}$, suggestive of an origin in
cloud cores of similar velocities.
The LSR velocities of the stars fall between 15.5 and 17.5~km~s\,$^{-1}$,
and those of the molecular cloud material in closest proximity are
$\sim10$~km~s\,$^{-1}$ (Figure 7).
It seems unlikely that these fairly young stars are unrelated,
however additional work remains in order to understand how the
stars are associated with each other and with the local molecular
material.  We discuss some suggestions for future research in Section 6.
One of the major mysteries in this region remains
the question of why there are not more young stars present.
We address this in Section 5.

%[scale=1.1,bb=5 15 90 160]
\begin{figure}[p]
%\plotone{prato_fig7.eps}
%\includegraphics[scale=0.73,bb=2 15 90 150]{prato_fig7.eps}
%\plotfiddle{prato_fig7.eps}{8.0in}{0.0}{72.0}{69.0}{-145.0}{6.0}
\centering
\includegraphics[width=0.7\textwidth,draft=False]{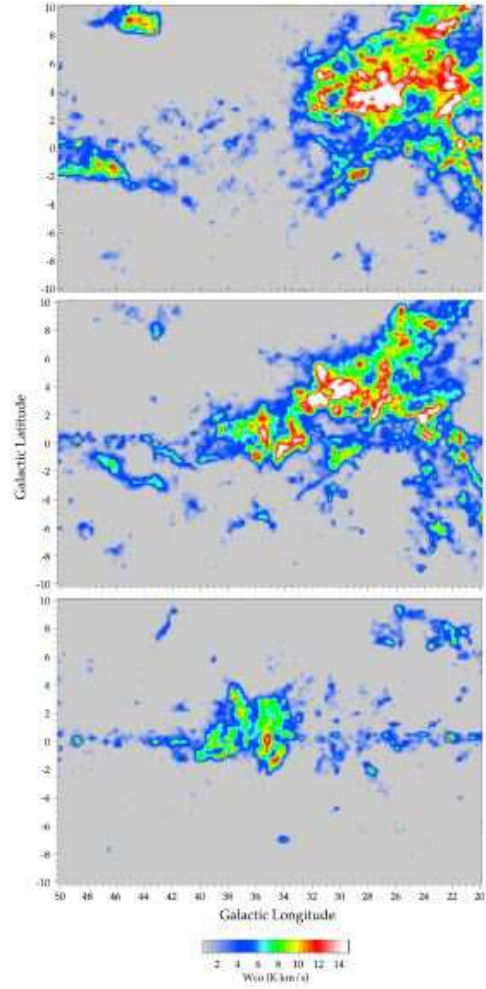}
\caption{$^{12}$CO maps integrated over 5~km~s\,$^{-1}$ intervals; the top one
centered at 5~km~s\,$^{-1}$, the middle one at
10~km~s\,$^{-1}$, and the bottom one at 15~km~s\,$^{-1}$.  Note the velocity break
at $l\sim33\,{\deg}$.  The young stars discussed in Section 3 are all
associated with molecular material with a $v_{lsr}\sim10$~km~s\,$^{-1}$
(see also Figure 4).}
\end{figure}

\subsection{Vulpecula and Scutum}

Further along the Galactic plane at longitudes  $l\sim55-63\,{\deg}$
lies the Vulpecula molecular cloud.  Most studies of Vulpecula
 have focussed on the open cluster in the cloud, {NGC 6823;}
at a distance of $\sim2-2.5$~kpc (Massey 1998)
only the higher mass stars in this cluster have been well studied
(e.g., Massey et al. 1995).  Kumar et al. (2004) estimate the age of
 {NGC~6823} at around 3~Myr, similar to the estimate
of Pe\~na et al. (2003) and consistent with the range,
2$-$7 Myr, determined by Massey (1998).  However, it is not clear
that the open cluster and the Vulpecula cloud are associated.
Dame \& Thaddeus (1985) note that there are two distinct distance
 scales for this region -- that of the OB1 association {(NGC 6823),}
2.3~kpc, and that
of the local dust, $\sim$400~pc (Neckel et al. 1980).  The
cluster is likely background to the molecular cloud.

Given the large difference in distance between either the dust or
the OB1 association in Vulpecula and the
stars and gas in the Aquila region, it is unlikely that
there is any direct relationship between these clouds.  Vulpecula falls
into an interstice between two Galactic spiral arms, roughly in the
direction of the Sun's motion into the first quadrant of the Galaxy
(Frisch 1998).  This provides a direct line of sight to the high-mass
cluster and recommends it for further study.

The Scutum region lies a couple degrees below the Galactic plane, just
 south of the {Aquila Rift} cloud in an apparent hole in the
molecular gas associated with the Rift.  The molecular gas velocities
of clouds in the direction of Scutum,
at 40$-$120~km~s\,$^{-1}$, are very distinct from those of
 the {Aquila Rift} (Dame et al. 1986). Santos et al. (2005) describe
the open cluster, M11 (NGC 6705), at a distance of 1.9~kpc, as
superimposed on the Scutum region.  Madsen \& Reynolds (2005) used
hydrogen emission lines to study the gas in the Scutum region.  They
conclude that this region is seen, at a distance of $\sim$6~kpc
with only $\sim$3 magnitudes of visual extinction,
through a gap in the more local molecular clouds, and is
associated with ionized gas in the inner Galaxy.

\section{Where Are All the Young Stars?}

Based on the very young ages of the observed stars in
Aquila, and the abundance of raw materials for star formation
present in the region, it is surprising that much larger numbers of
young sources are not found.  Why did the initial epoch of star
formation in this region produce so little?
 The Serpens star-forming region in the western part of the {Aquila
 Rift} is rich with hundreds of young stars (see chapter by
Eiroa et al.).  At Galactic longitude 33\,${\deg}$,
a distinct shift
in the velocity of the $^{12}$CO gas appears to demarcate a
boundary not only between Serpens and Aquila,
but between young star rich and poor regions, respectively (Figure 7).

Why is this the case?  Particularly striking is the eastern
 portion of the Rift that harbors only {HBC 684} as far as we know.
Possibly large
numbers of younger, embedded objects exist within the eastern
Rift but have so far evaded detection.  This bears further
attention, possibly in the form of Spitzer surveys, specifically of
the eastern Rift which appears to be gas-rich but young star poor.
Surveys for pre-stellar cores south of the eastern Rift
have not revealed a large population.

The small sample of young stars in the TW Hydrae Association
(e.g. Rich et al. 1999) are also spread out over an area of about
30~pc in diameter (Feigelson \& Montmerle 1999).
Aquila could comprise a similar sort of small association.
 {TW Hydrae} members have ages of $\sim$10~Myr; it is conceivable,
 even likely, that the {TW Hya}
stars formed in a much more compact region and have dispersed over
a large area over millions of years.  In the case of the sparse
population of Aquila stars, however, their $\sim$1~Myr ages imply
that a similar scenario is unlikely.  Furthermore, at least the
 partially embedded stars such as {AS 353} appear to be located close
to where they were formed.  Thus, the pre-main sequence
population in Aquila seems to be comprised of
numerous small pockets of star formation.

Frisch (1998) describes the properties of the local
interstellar medium within 500 pc of the Sun and points
 out that the {\it {``Aquila Rift} molecular cloud is the node region
where all of the superbubble shells from the three epochs of
 star formation in the {Scorpius-Centaurus Association,} as well as the
 most recent supernova explosion creating the {North Polar Spur,}
converged after plowing into the molecular gas and decelerating.''}
Could these dynamical processes in the Aquila region be
responsible for the disruption of star formation?  Or,
alternatively, could such processes have triggered only a very limited
epoch of star formation?  Without a much more exact study and
a comprehensive picture of the gas dynamics of the region,
it is not possible to evaluate these scenarios; however, they
remain intriguing possibilities.

A more mundane explanation for the dearth of young stars in the Aquila
molecular cloud may be simply that star formation has, so far, only
proceeded in isolated pockets.  If in the region of Aquila that lies
south of the Galactic plane the virial mass exceeded the total gas
mass, the complex may have broken up just before or while forming a
sporadic distribution of stars.  This is not a fully satisfactory
scenario, however, because ample raw materials abound in the eastern
Rift (Section 2).  Perhaps we have simply not yet determined the
absolute census of young Aquila objects; the location of these CO
clouds in the Galactic plane comes with contamination from a dense
stellar field within which it is highly non-trivial to pick out
T~Tauri stars (Figure 1).  Although this is a challenging undertaking,
a detailed survey for young stellar objects may be the most important
next step in furthering our understanding of star formation in Aquila
(Section 6).

\section{Future Observations: A Complete Census of Aquila Young Stars}

Objective prism and narrow-band H$\alpha$ imaging surveys
in Aquila should reveal emission line objects, typically
associated with accreting young stars.  Given the high disk
fraction (Section 3.5) among the known young candidate members,
this should reveal a significant portion of the pre-main sequence
stellar population, if indeed it is there.  A complementary approach
would be to use long-wavelength Spitzer observations to detect
directly the warm dust in the circumstellar disks of embedded protostars.

In the spirit of stellar characterization, it would be
profitable to understand the true space motions of the known
young Aquila stars (Ducourant et al. 2005).  By determining proper
motions and combining these with the recently measured radial velocities,
we will be able to determine the dynamical history of the
young stars with far more accuracy.

For those known pre-main sequence stars that are also
 sufficiently energetic radio sources (e.g., {AS 353),} very long
baseline interferometry can be used to determine very precise
stellar distances.  There currently exists an age-distance
degeneracy for the Aquila young stars, although the obvious
signatures of youth constrain them to be below a few Myr.
Accurate distance determinations would localize not only the
stars themselves, but also the associated molecular gas.

Millimeter wave surveys at higher angular
resolution using isotopes and species which trace the
denser regions of gas will provide a more complete picture
of the structure and dynamics of the Aquila clouds
and of how this region connects to the western part of the
 {Aquila Rift,} concurrent with the Serpens star-forming region.
Some preliminary observations of C$^{18}$O show that the
distribution of this tracer of denser gas is very different
from that of $^{12}$CO.  Cores revealed by such observations
may provide a guide for the most productive areas in which to
search for new or on-going star formation.

The Aquila region is rich, complex, and relatively nearby
and merits continued attention and study.

\bigskip
%\bigskip

{\bf
Acknowledgements}
B. Reipurth contributed to this chapter, particularly to Section 3.6; we
thank him for his assistance.
The authors are grateful to G. H. Herbig for interesting discussions
as well as for drawing
our attention to several important references.
We also thank L. Allen, C. Lada, M. Simon, and S. Strom for helpful
conversations about Aquila.  E. L. R. thanks I. S. McLean and L. P.
acknowledges Lowell Observatory for support during the
course of this project.  We are grateful to the referee, V. Strai\v{z}ys,
for his careful report, and to Bernhard Hubl for providing Figure~3.
This work made use of the SIMBAD reference database, the NASA
Astrophysics Data System, the NASA/IPAC Infrared Service Archive,
and the High Energy Astrophysics Science
Archive Research Center (HEASARC) at NASA/GSFC and the High Energy
Astrophysics Division of the Smithsonian Astrophysical Observatory.

\end{document}